\documentstyle[twocolumn,pre,aps]{revtex}

\begin{document}
\draft

\title{Combinatorial Approach to the Ground-State Energy of Square and
Triangular $ \pm J $ Spin Glasses}

\author{Pawe\l\ Polaszek \thanks{e-mail: polaszek@phys.amu.edu.pl}}
\address{Technische Universit\"{a}t Dresden, Institut f\"{u}r
Theoretische Physik, D-01062 Dresden, Germany}

\maketitle

\begin{abstract}
A new combinatorial, analytical approach to the ground-state energy problem
of spin glasses with different concentrations of $ \pm J $ interactions is
developed. The energy $ e_{0} $ is expressed in terms of the fraction of broken
bonds $ \mu_{0} $ and expanded into a fast converging series of the average
length of a
segment between two frustrated plaquettes of a minimum-weighted perfect
matching $ \overline{\Lambda} $. The concept of so called diagram stability
$ s $
is introduced in order to calculate coefficients of this expansion. Finally,
the fraction $ \mu_{0} $ as a function of the concentration $ p $ of negative
bonds is obtained for triangular and square infinite lattices in excellent
accordance with numerical simulations of large adequate systems.
\end{abstract}

\pacs{}

\section{Introduction} \label{intro}
The Ising model on a
regular lattice with $ \pm J $ interactions between nearest
neighbours is one of the simplest which is able to describe some
physical features of spin glasses appearing in nature.
Even for such "binary" interactions there is no analytical
approach to the ground-state energy apart from
unfrustrated Mattis spin glasses \cite{Mattis} which can be
transformed via gauge transformation into a ferromagnetic system.

Exact, numerical methods aimed at finding the ground-state energy can
only be applied to relatively small systems (as, e.g., the branch and bound
method \cite{branch,Grotschel}). Approximate methods are used for
larger systems (as, e.g., matching algorithms \cite{heu1,heu2}).
However, the numerical effort increases at least polynomially with system
size \cite{Barahona}. Thus the precision of an extrapolation to the
thermodynamic limit remains unknown. Therefore, any analytical
or semi-analytical approach to the problem is worth the effort. The
purpose of the present paper is to propose such an approach and to apply it
to the square and triangular lattices in order to find the fraction of
unsatisfied bonds in the ground state of an Ising system with randomly
mixed $ \pm J $ bonds.

The immediate stimulation has come from the
article of Liebmann and Schuster \cite{Schuster} who considered
the average minimal distance $ \overline{d}(p) $ between a frustrated
plaquette and the nearest other frustrated plaquette in the square
lattice. They used the following expansion:
\begin{equation}  \label{row:meand}
\overline{d}(p) = \sum_{i = 0}^{ \infty }C_{i}(p) \;\; ,
\end{equation}
where $ C_{i} $ is the conditional probability, namely the probability
that a plaquette is frustrated and all surrounding plaquettes up to
the distance $ d = i $ are not frustrated divided by the concentration
of frustrated plaquettes. The authors' main finding was that the parameter
$ \overline{d}(p) $ represent a continuous function of the concentration
$ p $ of negative bonds. They stated that series (\ref{row:meand}) converged
rapidly and found that an excellent approximation for $ \overline{d}(p) $
was obtained using only the first three (or even two) terms. This allows one
to hope that a similar approximation should also work for an analogous series
expansion of the energy.

The parameter $ \overline{d}(p) $ was also investigated by Bendisch
\cite{Bendisch}.
He found that the maximum of $ \overline{d}(p) $ at $ p \approx 0.10 $
($ p \approx 0.15 $) for the square (triangular) lattice reflects
an inflection point in the ground-state magnetisation.

In the present work, a new method of combinatorial approach to
a minimum-weighted perfect matching of the graph of frustrated plaquettes
has been developed in order to obtain the expansion of the ground-state
energy as a function of the concentration of negative bonds.

\section{Basic definitions}

The Ising Hamiltonian without external magnetic field is given by:
\begin{equation} \label{def:ham}
H = - \sum_{i,j}J_{ij}S_{i}S_{j} \;\; .
\end{equation}
The interaction between the $i$-th and $j$-th spins has a ferromagnetic
character for $ J_{ij} > 0 $ and otherwise an antiferromagnetic one. We will
use the term of a positive (negative) bond interchangeably with the
term of a ferromagnetic (antiferromagnetic) interaction. We restrict
ourselves to the following probability distribution of
$ J_{ij} $:
\begin{equation} \label{def:distr}
f(J_{ij}) = (1-p)\delta (J_{ij}-1) + p\delta(J_{ij}+1) \;\;
\end{equation}
for nearest neighbours (otherwise $ f(J_{ij}) = 0 $).
The parameter $ p \in [0,1] $ has a clear interpretation as the
concentration of negative bonds or the probability that an interaction
is antiferromagnetic. We will consider the two dimensional triangular and
square lattices (coordination numbers 6 and 4, respectively).

The unit cell of the 2D lattice will be called a plaquette.
The term of frustration is reserved for systems: the smallest
system is a plaquette, so we will say that a plaquette is frustrated
(unfrustrated) when an odd (even) number of plaquette bonds is
negative. This term is not adequate for bonds because there
always exist two configurations of connected spins
which locally minimize the energy. Thus, we will say that a bond is
unsatisfied or broken when the configuration of connected spins
does not minimize the energy locally.

A useful formulation of the ground-state problem can be obtained
using the duality concept \cite{Barahona}, \cite{Bieche}, as
a problem of finding a minimum-weighted perfect matching in the graph
of frustrated plaquettes. Thus, the ground-state energy can be expressed as:
\begin{equation} \label{row:energy}
E_{0} = - \frac{Nc}{2} + 2w_{0} \;\; ,
\end{equation}
where $ N $ is the number of spins, $ c $ is the coordination number
and $ w_{0} $ is the weight of a minimum-weighted perfect matching which
equals to the total number of broken bonds in the real lattice.
The ground-state energy per bond $ e_{0} $ reads:
\begin{equation} \label{row:intensive}
e_{0} = -1 + 2\mu_{0} \;\; ,
\end{equation}
where the parameter
\begin{equation}
\mu_{0} = \frac{w_{0}}{Nc/2}
\end{equation}
has a clear interpretation as the fraction of bonds unsatisfied in a ground
state. The parameter $ \mu_{0} $ admits an interesting generalization
in \cite{misfit} where the so called "misfit parameter"
is defined as:
\begin{equation}  \label{def:misfit}
\mu_{0} = \frac{E_{0} - E^{id}_{min}}{E^{id}_{max} - E^{id}_{min}} \;\; .
\end{equation}
$ E^{id}_{min} $ and $ E^{id}_{max} $ are minimal and maximal energies,
respectively, of an adequate "ideal" system which is analogous to the
original one but it is not frustrated. The advantage of definition
(\ref{def:misfit}) is that it allows a comparison of frustrated systems
of different nature. In our case, the "ideal" system is a Mattis spin glass,
thus $ E^{id}_{min} = - \frac{Nc}{2} = - E^{id}_{max} $
and simple algebra leads again to formula (\ref{row:intensive}).

The notion of the fraction of bonds broken in a ground state
appeared in the literature earlier \cite{Stein}. For convenience
justified by expression (\ref{row:intensive}) for $ e_{0} $ as
a linear function of $ \mu_{0} $, from now on we will be concerned with
calculating
the fraction $ \mu_{0} $, i.e., the misfit parameter. Also final
results will be presented in terms of this quantity.

\section{Series expansion of the misfit parameter}
The weight of a perfect matching in a ground state is equal to the total sum
of lengths of segments each of them connecting a pair of frustrated
plaquettes:
\begin{equation} \label{row:weight}
w_{0} = \sum_{\Lambda = 1}^{\infty }\Lambda N_{ \Lambda } \;\; ,
\end{equation}
where $ N_{ \Lambda }$ is the number of segments of length
$ \Lambda $ or
\begin{equation}
w_{0} = \frac{N_{f}}{2}\overline{\Lambda } \;\; ,
\end{equation}
with $ N_{f} $ as the number of frustrated plaquettes and
$ \overline{ \Lambda } $ as the average segment length:
\begin{equation}  \label{def:meanseg}
\overline{ \Lambda } = \sum_{\Lambda = 1}^{\infty}\Lambda P( \Lambda ) \;\; .
\end{equation}
$ P( \Lambda ) $ is the probability that a randomly chosen segment has
a length $ \Lambda $. Clearly:
\begin{equation}
P( \Lambda ) = \frac{N_{ \Lambda }}{\sum_{ \Lambda ' = 1 }^{ \infty }
N_{\Lambda '}}
\end{equation}
and $ N_{f}/2 = \sum N_{ \Lambda } $, so formula (\ref{row:weight})
holds. The misfit parameter can be written:
\begin{equation}
\mu_{0} = \frac{N_{f}}{Nc} \overline{ \Lambda }  \;\; .
\end{equation}
Elementary, geometrical considerations lead to a relation between
the coordination number and the total number of plaquettes in a regular
lattice (neglecting boundary effects):
\begin{equation}
N_{f} = \frac{N(c-2)}{2}P_{F}(p) \;\; ,
\end{equation}
where $ P_{F}(p) $ is the probability that a plaquette is frustrated in
the system with a concentration $ p $ of negative bonds. From now on,
we will omit in the notation $ P_{F}(p) $ the explicit dependence on $ p $,
and we will write just $ P_{F} $. Finally, we obtain a useful formula:
\begin{equation}  \label{row:main}
\mu_{0} = \frac{c-2}{2c} \overline{ \Lambda } P_{F} \;\; .
\end{equation}

At this point, we stress that the statistics $ P( \Lambda ) $ depends on the
choice of one of the ground states. For example, in Fig.\ \ref{diffstat} are
presented two possible minimum-weighted perfect matchings: one of them
provides two segments of length 2 and the other one gives one segment of
length 1 and one segment of length 3. Thus, there is no universal
distribution of $ P( \Lambda ) $, but this distribution may vary from one
ground state to another. The common feature of these distributions is that
they must give the same mean length of segments $ \overline{ \Lambda } $.
To keep symbols as simple as possible we will not introduce a new index
like $ P_{m}( \Lambda ) $, but the reader should keep in mind that
calculations are performed for a specific ground state. The choice of this
ground state will be discussed later.

\section{Method}
Referring to the discussion in section\ \ref{intro}\ we
restrict ourselves to segments with lengths 1, 2 or 3.
Hence, approximately, $ P(\Lambda \geq 4) = 0 $. It follows:
\begin{equation}
\overline{\Lambda} = \frac{3P(3)+2P(2)+P(1)}
{P(3)+P(2)+P(1)}
\end{equation}
or
\begin{equation} \label{row:lambda}
\overline{\Lambda} = \frac{3R_{3,2}R_{2,1}+2R_{2,1}+1}
{R_{3,2}R_{2,1}+R_{2,1}+1} \;\; ,
\end{equation}
where $ R_{i,j} = P(i)/P(j) $.

To perform calculations we choose a specific ground state, namely a state
with as few as possible matching intersections. Fig.\ \ref{choice}
illustrates the rule of thumb for such a choice (sometimes it is not possible
to avoid intersections - Fig.\ \ref{choice}c). Next we perform a
transformation of the system such that in the transformed system only broken
bonds are negative  (Fig.\ \ref{gauge}). This transformation is defined as
the mapping: $ J_{ij} \rightarrow |J_{ij}| $ for satisfied bonds and
$ J_{ij} \rightarrow -|J_{ij}| $ for broken bonds. This mapping does not
change the configuration of frustration of the system. Thus, both systems are
equivalent with respect to the gauge symmetry in the meaning of
\cite{Balian}. In particular, they have the same ground-state energy.
Furthermore, in the transformed system the concentration of negative bonds is
equal to the fraction $ \mu_{0} $ of bonds broken in the ground state, i.e.,
to the misfit parameter.

\subsection{ $ R_{2,1} $ for the triangular lattice }
In Fig.\ \ref{trdiag}, we show the three diagrams which contain segments of
length 1 and the six diagrams with segments of length 2. In the present
approximation, diagrams with matching intersections, such as shown
in Fig.\ \ref{choice}c are omitted. This is justified by our choice of the
ground state with as few intersections as possible. Because of the rotational
symmetry between diagrams considered one can write:
\begin{equation}  \label{def:r21}
R_{2,1} = \frac{6}{3}\frac{n_{2a}}{n_{1a}} =
2\frac{n_{2a}}{n_{1a}} \;\; ,
\end{equation}
where $ n_{1a} $ and $ n_{2a} $ are the probabilities of finding diagrams
of type 1a and 2a, respectively, in the lattice.

To calculate the ratio $ n_{2a}/n_{1a} $ we add a plaquette
to the diagram 1a in order to have the same number of plaquettes and bonds as
in the diagram 2a. This additional plaquette may be either frustrated or not
and we have three new diagrams: 1a', 1a'', 1a''$\!$' (Fig.\ \ref{trnew}).
If the lattice were a set of isolated diagrams (which is not the case) like
diagrams 1a', 1a'', 1a''$\!$' and 2a, we could write:
\begin{eqnarray} \label{row:ind}
\frac{n_{2a}}{n_{1a'}} & = & \frac{P^{2}_{F}(1-P_{F})\mu^{2}_{0}
(1-\mu_{0})^{5}}{P^{2}_{f}(1-P_{F})\mu_{0}(1-\mu_{0})^{6}} =
\frac{\mu_{0}}{1-\mu_{0}} \;\; , \nonumber \\
\frac{n_{2a}}{n_{1a''}} & = & \frac{P^{2}_{F}(1-P_{F})\mu^{2}_{0}
(1-\mu_{0})^{5}}{P^{3}_{F}\mu^{2}_{0}(1-\mu_{0})^{5}} =
\frac{1-P_{F}}{P_{F}} \;\; , \nonumber \\
\frac{n_{2a}}{n_{1a'''}} & = & \frac{P^{2}_{F}(1-P_{F})\mu^{2}_{0}
(1-\mu_{0})^{5}}{P^{2}_{F}(1-P_{F})\mu^{3}_{0}(1-\mu_{0})^{4}} =
\frac{1-\mu_{0}}{\mu_{0}} \;\; .
\end{eqnarray}
The equations above contain information on isolated diagrams.
Diagrams 1a' (or 1a''$\!$') and 2a have the same number of frustrated and
unfrustrated plaquettes but different numbers of satisfied and broken
bonds. Diagrams 1a'' and 2a have the same number of satisfied and broken
bonds but differ in the numbers of frustrated and unfrustrated plaquettes.
All these diagrams have the same total number of plaquettes and bonds.
Hence, for isolated diagrams (i.\ e., independent of any influence of the
outer part of the infinite lattice) (\ref{row:ind}) would hold.
However, the diagrams are situated inside the lattice and feel the presence
of other frustrations. Therefore, for diagrams within the lattice, we
introduce a new parameter, called stability $ s_{\alpha} $ of diagram
$ \alpha $, such that:
\begin{eqnarray} \label{row:stab}
\frac{n_{2a}}{n_{1a'}} & = & \frac{\mu_{0} s_{2a}}{(1-\mu_{0})s_{1a'}}
\;\; ,  \nonumber \\
\frac{n_{2a}}{n_{1a''}} & = &
\frac{(1-P_{F})s_{2a}}{P_{F}s_{1a''}} \;\; , \nonumber \\
\frac{n_{2a}}{n_{1a'''}} & = &
\frac{(1-\mu_{0})s_{2a}}{\mu_{0} s_{1a'''}} \;\; .
\end{eqnarray}

What is the meaning of the stability parameter? It measures the probability
of the occurrence of such a frustration in the outer part of the lattice
which destroys the matching of a diagram, i.e., some of the segments vanish
and others appear. Fig.\ \ref{destr} shows an example of a destruction of the
matching in diagram 2a. Destruction of the matching appears if a loop exists
containing a given diagram and for which an alternative matching is
energetically more profitable. Accordingly, the stability parameter can be
expressed as a loop series:
\begin{equation}  \label{row:loop}
s_{\rm diag.} = 1 - \sum_{\rm loops}D_{loop} \;\; ,
\end{equation}
where the sum goes over all loops containing a given diagram and
$ D_{loop} $ is the probability that there appears a frustration
destroying the matching of a diagram inside a loop.

Because a segment of length 1 is energetically the most profitable one, we
admit:
\begin{equation}  \label{row:stab1a}
s_{1a'} = s_{1a''} = s_{1a'''} \approx 1 \;\; .
\end{equation}
For $ s_{2a} $ we will consider only the first order loops in the expansion
(\ref{row:loop}). We are left with one type of smallest loops with six
plaquettes (Fig.\ \ref{destr}). Frustration of plaquettes belonging to the
diagram is fixed. Furthermore there are only two configurations of
frustration of the remaining plaquettes which destroy the matching of diagram
2a. One of these configurations contains two frustrated and one unfrustrated
plaquette and the other contains three frustrated plaquettes. Thus, the
contribution to the stability $ s_{2a} $ up to first order is:
\begin{equation}
D_{6} = \frac{1}{3}(1-P_{F})p^{2}_{F} + p^{3}_{F} =
\frac{1}{3}p^{2}_{F} + \frac{2}{3}p^{3}_{F}
\end{equation}
and
\begin{equation} \label{row:stab2a}
s_{2a} \approx 1 - \frac{1}{3}p^{2}_{F} - \frac{2}{3}p^{3}_{F} \;\; .
\end{equation}
Combining (\ref{def:r21}), (\ref{row:stab}) and (\ref{row:stab1a}) we obtain:
\begin{eqnarray} \label{row:r21}
R_{2,1} & = & 2\left(\frac
{{\rm n_{1a'}}+2\;{n_{1a''}}+{n_{1a'''}}}{n_{2a}}
\right)^{-1} =  \nonumber \\
 & = & 2\left(\frac{1-\mu_{0}}{\mu_{0}s_{2a}}+2\frac{P_{F}}{(1-P_{F})s_{2a}}+
\frac{\mu_{0}}{(1-\mu_{0})s_{2a}}\right)^{-1} = \\
 & = & \frac{2\mu_{0}(1-\mu_{0})(1-P_{F})s_{2a}}
{(1-\mu_{0})^{2}(1-P_{F})+2\mu_{0}(1-\mu_{0})P_{F}+\mu_{0}^{2}(1-P_{F})} \;\; ,
\nonumber
\end{eqnarray}
with $ s_{2a} $ given by (\ref{row:stab2a}).

\subsection{ $ R_{3,2} $ for the triangular lattice}
There are six diagrams which contain segments of length 2 and
twelve diagrams with segments of length 3 (Fig.\ \ref{trdiag}). As before,
we omit
diagrams with matching intersections. A practical approximation is made:
\begin{equation}  \label{row:r32}
R_{3,2} \approx \frac{12}{6}\frac{n_{3a}}{n_{2a}} \approx
2\frac{n_{2a}}{n_{1a}} = R_{2,1} \;\; .
\end{equation}

\subsection{Final formula for the triangular lattice}
Using (\ref{row:main}), (\ref{row:lambda}) and (\ref{row:r32}) we get:
\begin{equation}  \label{row:trfinal}
2c\mu_{0} - (c-2)P_{F}\frac{3R^{2}_{2,1}+2R_{2,1}+1}
{R^{2}_{2,1}+R_{2,1}+1} = 0 \;\; ,
\end{equation}
where for the triangular lattice the probability that the plaquette is
frustrated amounts to:
\begin{equation} \label{row:frtr}
P_{F} = p(4p^{2}-6p+3) \;\; .
\end{equation}
The last two equations combined with (\ref{row:r21}) for $ R_{2,1} $ and
(\ref{row:stab2a}) for $ s_{2a} $ make it possible to calculate the misfit
parameter $ \mu_{0} $. To solve this system of equations numerically
a simple bisection algorithm was applied.
Results are presented in Fig.\ \ref{results}a together with the
data of simulations of large samples \cite{Bendisch2,Vogel}.

\subsection{$ R_{2,1} $ for the square lattice}
As in the triangular lattice, we choose
the ground state with as few matching intersections as possible
and again we perform a gauge transformation such that
in the transformed system only broken bonds are negative.

There are two diagrams which contain segments of length 1 and six
diagrams which contain segments of length 2 (Fig.\ \ref{sqdiag}).
As before, we omit diagrams with matching intersections in accordance with
the foregoing choice of the ground state. Thus, for the square lattice we
obtain
a formula analogous to (\ref{def:r21}):
\begin{equation} \label{def:sqr21}
R_{2,1} = \frac{4}{2}\frac{n_{2a}}{n_{1a}} + \frac{2}{2}\frac{n_{2e}}{n_{1a}}
\approx
3\frac{n_{2a}}{n_{1a}} \;\; .
\end{equation}
However, not all segments of length 2 have the same symmetry in the square
lattice (Fig.\ \ref{sqdiag}  diagrams 2a and 2e), so unlike (\ref{def:r21}),
eq. (\ref{def:sqr21}) is only an approximation.
We add a plaquette to the diagram 1a to have the same number of plaquettes
and bonds as in diagram 2a. We have three new diagrams: 1a', 1a'' and
1a''$\!$' (Fig.\ \ref{sqnew}). Applying the concept of the diagram stability
for the square lattice we get an equation identical to (\ref{row:stab}).
As before, the approximation (\ref{row:stab1a}) is made, since segments of
length 1 are energetically the most profitable.

Now, the first-order loop expansion (\ref{row:loop}) for $ s_{2a} $ for the
square
lattice  will be considered. We have two smallest loops with six
plaquettes containing diagram 2a (Fig.\ \ref{sqdestr}).
The contributions from these two loops are:
\begin{equation}
D_{6} \approx \frac{2}{3}(1-P_{F})p^{2}_{F} + 2p^{3}_{F} =
\frac{2}{3}p^{2}_{F} + \frac{4}{3}p^{3}_{F} \;\; .
\end{equation}
Thus, for the square lattice we have:
\begin{equation} \label{row:sqstab2a}
s_{2a} \approx 1 - \frac{2}{3}p^{2}_{F} - \frac{4}{3}p^{3}_{F} \;\; .
\end{equation}
Formally, $ 0 \leq P_{F} \leq 1 $ as $ P_{F} $ is a probability but
for the square lattice $ P_{F} \leq 0.5 $, so there is no danger
that thr above approximation leads to an unphysical, negative stability.
Combining (\ref{def:sqr21}), (\ref{row:stab}) and (\ref{row:stab1a})
we obtain:
\begin{eqnarray} \label{row:sqr21}
R_{2,1} & = & 3\left(\frac{n_{1a'}+3n_{1a''}+3n_{1a'''}}{n_{2a}}\right)^{-1} =
\nonumber \\
 & = & 3\left(\frac{1-\mu_{0}}{\mu_{0}s_{2a}}+3\frac{P_{F}}{(1-P_{F})s_{2a}}+
3\frac{\mu_{0}}{(1-\mu_{0})s_{2a}}\right)^{-1} =  \\
 & = & \frac{3\mu_{0}(1-\mu_{0})(1-P_{F})s_{2a}}
{(1-\mu_{0})^{2}(1-P_{F})+3\mu_{0}(1-\mu_{0})P_{F}+3\mu_{0}^{2}(1-P_{F})} \;\;
, \nonumber
\end{eqnarray}
with $ s_{2a} $ given by (\ref{row:sqstab2a}).

\subsection{ $ R_{3,2} $ for the square lattice }
There are six diagrams which contain segments of length 2 and
fourteen diagrams with segments of length 3 (Fig.\ \ref{sqdiag}). We omit
diagrams with matching intersections, as before, and make a practical
approximation is made:
\begin{equation}  \label{row:sqr32}
R_{3,2} \approx \frac{14}{6}\frac{n_{3a}}{n_{2a}} \approx
\frac{7}{3}\frac{n_{2a}}{n_{1a}} = \frac{7}{9}R_{2,1} \;\; .
\end{equation}

\subsection{Final formula for the square lattice}
Combining (\ref{row:main}), (\ref{row:lambda}) and (\ref{row:sqr32}) we get:
\begin{equation}  \label{row:sqfinal}
2c\mu_{0} - (c-2)P_{F}\frac{\frac{7}{3}R^{2}_{2,1}+2R_{2,1}+1}
{\frac{7}{9}R^{2}_{2,1}+R_{2,1}+1} = 0 \;\; ,
\end{equation}
where the probability that the plaquette is frustrated in the square lattice
is:
\begin{equation} \label{row:frsq}
P_{F} = 4p(1-p)\left(p^{2}+(1-p)^{2}\right) \;\; .
\end{equation}
The last two equations when combined with (\ref{row:sqr21}) for $ R_{2,1} $
and (\ref{row:sqstab2a}) for $ s_{2a} $ make it possible to calculate the
misfit parameter $ \mu_{0} $ numerically. Results are presented in
Fig.\ \ref{results}b together with the data of simulations of large
samples \cite{Bendisch2,Vogel}.

\section{Conclusions}
The ground-state energy $ e_{0} $ per bond is a simple linear function
(\ref{row:intensive}) of the misfit parameter $ \mu_{0} $ and
the present approach provides values of the misfit parameter in the ground
state of mixed Ising ferro-antiferro magnets in excellent agreement with
numerical experiments \cite{Bendisch2} for large
triangular and square lattices. This proves the effectiveness of the method
and justifies {\it a posteriori} the approximations made. The results for
the boundary values of the concentration $ p $ of negative bonds, e.g., for
$ p = 0 $ or $ p = 1 $ are equal to the well known exact solutions in these
cases. In fact, for the square lattice in both cases $ P_{F} = 0 $ and
because $ R_{2,1} $ given by (\ref{row:sqr21}) has no singularities the only
solution of (\ref{row:sqfinal}) is $ \mu_{0} = 0 $.

The situation for the triangular lattice is the same for $ p = 0 $. On the
contrary, for $ p  $ tending to unity $ P_{F} $ tends to 1 and
(\ref{row:stab2a}) with (\ref{row:r21}) lead to $ R_{2,1} \rightarrow 0 $ and
$ \mu_{0} \rightarrow 1/3 $. The plaquettes in the triangular lattice contain
an odd number of bonds. Thus, in the antiferromagnetic limit the system
becomes a fully frustrated one and elementary considerations prove that
indeed one third of the bonds must be broken in any ground state
($ \mu_{0} = 1/3 $).

The case of $ p = 0.5 $ is under particular considerations of many
authors. Thus, the solutions for this concentration of negative bonds are
specified: $ \mu_{0}^{tr} =0.2218 $ and $ \mu_{0}^{sq} =0.1546 $.

Each step of the foregoing approximations can be developed relatively easily
up to the second and further orders of precision. The method is open to such
improvements. At this point, one of possible directions of the development
should be mentioned which is somewhat less evident. We approximated the
probability of finding a diagram or a loop by the formula
$ (1-P_{F})^{N_{u}}p^{N_{f}}_{F} $, where $ N_{f} $ is the number
of frustrated plaquettes and $ N_{u} $ is the number of those unfrustrated
plaquettes out of which a given diagram is built. It is not quite correct and
hopefully could be replaced by an exact formula in future calculations.

The method can be applied to any two dimensional regular lattice.
An extension on the case of three dimensional cubic lattice is under the
consideration.

\section{Acknowledgements}
The author would like to thank Dresden research group: S. Kobe, T. Klotz
and J. Wei\ss barth for fruitful discussions, advices and critical reading
of the manuscript as well as for the warm hospitality extended to him when
in Dresden. Special thanks are due to J. Bendisch for the permission of
comparison with and of publishing the results of his numerical simulations.
The author acknowledges also A. R. Ferchmin for the critical reading of
the manuscript.

The work was supported by the DFG (project no. Ko 1416).

\pagebreak

\pagebreak

\begin{figure}
\caption{Two different ground states contributing to different
probability distributions $ P(\Lambda) $.} \label{diffstat}
\end{figure}

\begin{figure}
\caption{a) One specific ground state without a matching intersection of
segments representing the basis of the method. b) A ground state with
a matching intersection that can be transformed to a) via an inversion
of the proper spin (marked with $ \downarrow $).
c) A configuration of frustrated plaquettes necessitating intersection
(disregarded in this work).} \label{choice}
\end{figure}

\begin{figure}
\caption{The gauge transformation bound to a given ground state leading from
the initial lattice (a) to the lattice with only broken bonds being
negative (b).} \label{gauge}
\end{figure}

\begin{figure}
\caption{Diagrams in the triangular lattice taken into account
during considerations. Diagrams with matching intersections were not
considered.} \label{trdiag}
\end{figure}

\begin{figure}
\caption{Diagrams 1a', 1a'', 1a''$\!$' and 2a taken into account in the
calculation of the parameter $ R_{2,1} $ of the triangular lattice. Note
that there are two possibilities for diagram 1a''.}
\label{trnew}
\end{figure}

\begin{figure}
\caption{Two frustration configurations inside the smallest loop destroying
the matching of diagram 2a of the triangular lattice. Solid lines represent
new segments, crossed lines - destroyed segments.} \label{destr}
\end{figure}

\begin{figure}
\caption{Diagrams in the square lattice taken into account.
Diagrams with matching intersections were not considered.} \label{sqdiag}
\end{figure}

\begin{figure}
\caption{Diagrams 1a', 1a'', 1a''$\!$' and 2a taken into account in the
calculation of the parameter $ R_{2,1} $ of the square lattice. Note that
there are three possibilities for diagram 1a'' and also three possibilities
for diagram 1a''$\!$'.} \label{sqnew}
\end{figure}

\begin{figure}
\caption{The two smallest loops containing the diagram 2a in the square
lattice which contribute to the stability calculation.} \label{sqdestr}
\end{figure}

\begin{figure}
\caption{ Misfit parameter $ \mu_{0} $ as a function of the concentration
$ p $ of negative bonds for the triangular lattice (a) and for the square
lattice (b), obtained in this work. Crosses denote results obtained by
numerical simulations of large samples (up to 200 spins) [13].
Rhombs denote average values over 500 samples for each concentration of
negative bonds for square and triangular lattices of size $ 6 \times 8 $ and
$ 6 \times 6 $, respectively [14].}
\label{results}
\end{figure}


\begin{references}

\bibitem[1]{Mattis}
D. C. Mattis, Phys. Lett. A {\bf 56} 421 (1976).

\bibitem[2]{branch}
S. Kobe and A. Hartwig, Comp. Phys. Comm. {\bf 16} 1 (1978);
A. Hartwig, F. Daske and S. Kobe, Comp. Phys. Comm. {\bf 32} 133 (1984).

\bibitem[3]{Grotschel}
M. Gr\"{o}tschel, M. J\"{u}nger and G. Reinelt, {\it Heidelberg Colloquium
on Glassy Dynamics} edited by J. L. van Hemmen and I. Morgenstern,
Lecture Notes in Physics {\bf 275} 325 (1987).

\bibitem[4]{heu1}
M. Achilles, J. Bendisch, K. Cassirer and H. von Trotha, {\it Gesellschaft
f\"{u}r Mathematik und Datenverarbeitung}, D-5205 St. Augustin, Germany,
GMD-Studies {\bf 186} (1991).

\bibitem[5]{heu2}
H. Freund and P. Grassberger, J. Phys. A: Math. Gen. {\bf 22} 4045 (1989).

\bibitem[6]{Barahona}
F. Barahona, J. Phys. A: Math. Gen. {\bf 15} 3241 (1982).

\bibitem[7]{Schuster}
R. Liebmann and H. G. Schuster, J. Phys. C: Solid State Phys. {\bf 14} 709
(1981).

\bibitem[8]{Bendisch}
J. Bendisch, Phys. A {\bf 202} 48 (1994).

\bibitem[9]{Bieche}
I. Bieche, R. Maynard, R. Rammal and J. P. Uhry, J. Phys. A: Math. Gen.
{\bf 13} 2553 (1980).

\bibitem[10]{misfit}
S. Kobe and T. Klotz, Phys. Rev. E {\bf 52} (1995).

\bibitem[11]{Stein}
D. L. Stein, G. Baskaran, S. Liang and M. N. Barber,
Phys. Rev. B {\bf 36} 5567 (1987).

\bibitem[12]{Balian}
R. Balian, J. M. Drouffe and C. Itzykson, Phys. Rev. D {\bf 11} 2098 (1975).

\bibitem[13]{Bendisch2}
J. Bendisch, Unpublished data obtained from numerical simulations
of up to 200 spins (matching algorithm), private communication.

\bibitem[14]{Vogel}
W. Lebrecht and E. E. Vogel, Average values of 500 samples for each
concentration of negative bonds for square and triangular lattices of
size $ 6 \times 8 $ and $ 6 \times 6 $, respectively (by enumerating the
lower-energy portion of the configuration space), private communication.

\end{references}
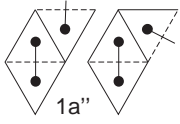
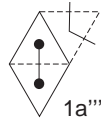
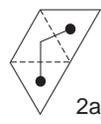
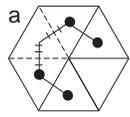
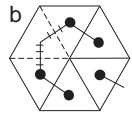
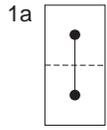
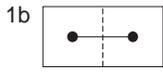
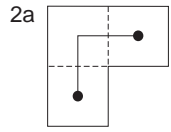
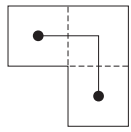
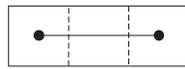
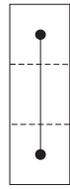
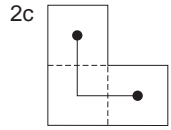
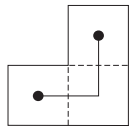
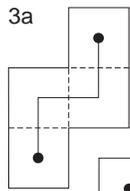
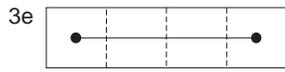
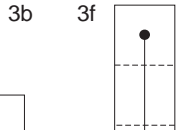
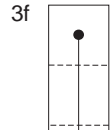
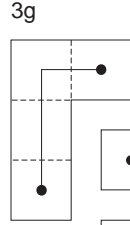
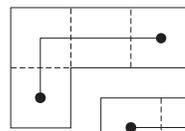
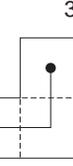
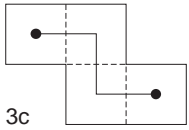
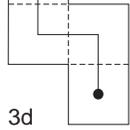
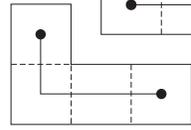
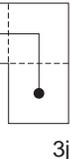
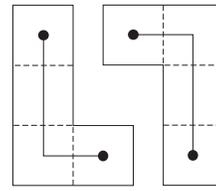
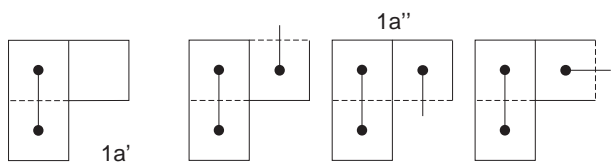
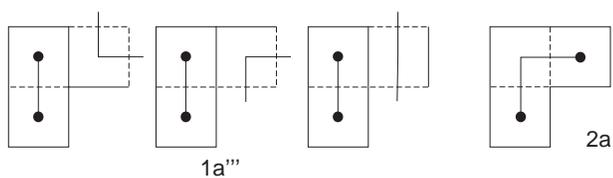
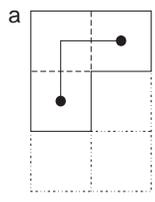
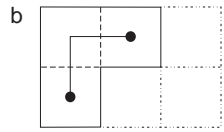
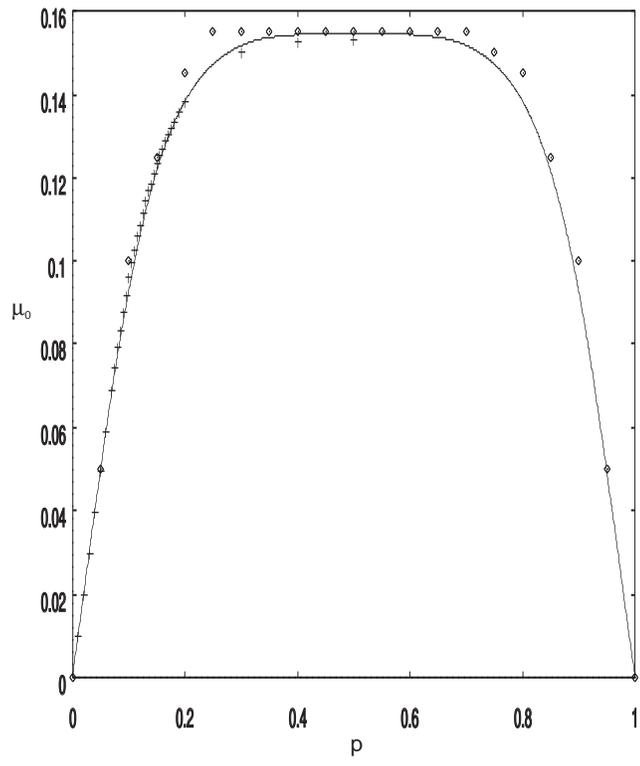
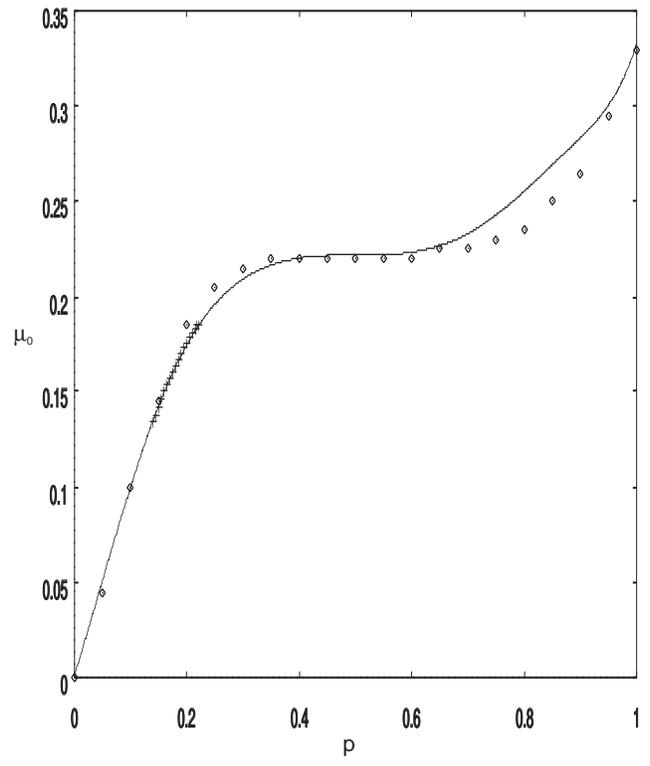
\end{document}